\documentstyle[11pt,twoside,colloqOHP,epsfig]{article}
\def\ms{m$\,$s$^{-1}$}
\begin{document}
\title{Photometric searches for transiting planets:\\ results and challenges}
\author{Fr\'ed\'eric Pont}
\affil{Geneva Observatory [frederic.pont@obs.unige.ch]} 
%\affil{$^2$Institute address [e-mail]} 
%
\setcounter{secnumdepth}{1}
\begin{abstract}
Ground-based photometric surveys have led to the discovery of six transiting exoplanets, five of which were detected by the OGLE surveys. The FLAMES multi-object spectrograph on the VLT has permitted a very efficient follow-up of the OGLE transit candidates, characterising not only the 5 planets but also more than 50 systems producing similar photometric signatures -- mainly eclipsing binaires. The presence of these ubiquitous ``impostors'' is a challenge for transit surveys. Another difficulty is the presence of red noise in the photometry, which implies a much lower sensitivity to transiting planets than usually assumed. We outline a method to estimate how the red noise will affect the expected yield of photometric transit searches.
\end{abstract}

\section{Introduction}

\begin{figure}[htb]
\epsfig{file=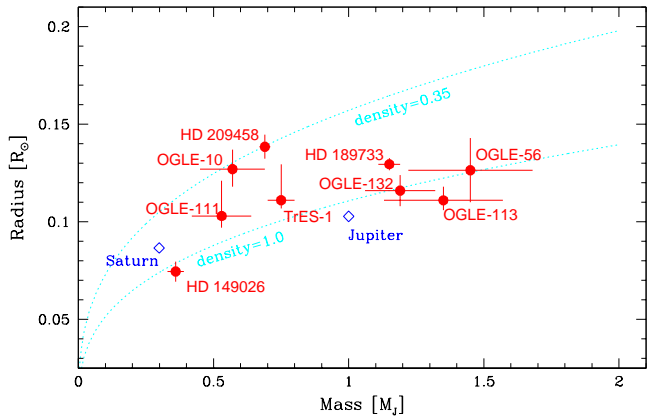,width=9cm}
\centering
%\plotone{box.ps}
\caption{Mass-radius relation for the known transiting exoplanets.}
\label{MMR}
\end{figure}

 During the first 8 years after the discovery of 51 Pegb, only one transiting exoplanets was found. But now, as we mark the 10th anniversary, nine are known (Fig.~\ref{MMR}). Three of them were discovered by radial velocity planet searches, and six by ground-based photometric transit surveys. 
 
The statistics of gas giant exoplanets inferred from radial velocity surveys indicate that about one star in a thousand should be transited by a hot Jupiter. Assuming a solar-size star and a Jupiter-size planets, such transits produce periodic 1\% dips in the ligth curve of the star lasting 2-4 hours. Hence the idea of detecting transiting gas giants by monitoring in photometry a few thousand stars with an accuracy of better than 1\%. 
Since the confirmation of the first transiting exoplanet by Charbonneau et al. (2000), several dozen photometric transit surveys were started, from the ground with small telescopes (10-20cm) on wide fields and relatively bright stars (10-14 mag), with larger telescopes (1-2m) on smaller fields and fainter stars (14-20 mag), and from space with the HST. On paper, these surveys could be expected to detect dozens or even hundreds of hot Jupiters (e.g. Horne 2001). Even the most modest of the surveys were expected to find one or two transiting hot Jupiters each season. 

The results up to now, however, have been meager in comparison with initial expectations. Most surveys have failed to confirm any transiting exoplanet candidate. Indeed only two surveys, the TrES network (Alonso et al. 2004) and the OGLE survey (Konacki et al. 2003, Bouchy et al. 2004, Pont et al. 2004, Konacki et al. 2005) have yielded any detection at all. The OGLE survey alone can be credited with 5 of the 6 transiting planets found by ground-based transit surveys.  In this review, I will concentrate on the OGLE survey, considering  how the conclusions from the analysis and follow-up of the OGLE data are also relevant to other surveys generally, in particular to the issue of why the yield of most surveys has been modest.

\section{The OGLE planetary transit candidates}
Using the 1.3-m Warsaw University Telescope at Las Campanas Observatory (Chile) with an 8k$\times$8k CCD mosaic covering a 0.34 deg$^{2}$ field-of-view, 
the OGLE-III survey (Udalski et al. 2002a-c, 2004) has realised an extensive photometric search for planetary and low-luminosity object transits. 
In four observing seasons, about 3 square degrees near the Galactic plane were monitored for periodic 
eclipse signals with depth from a few per cent down to slightly
below one per cent. Altogether 177 shallow periodic transit signals  were detected and announced . 
The radii of the transiting low-luminosity objects, estimated from the shape of the transit signal, 
range from 0.5 Jupiter radius to 0.5 solar 
radius, and their orbital periods from 0.8 to 8 days. The smallest objects 
could be suspected to be extrasolar giant planets, but the radius 
estimated from the photometric signal is not sufficient to conclude on the planetary nature of the objects. 
They could as well be brown dwarfs or low-mass stars,  since in the low mass regime ($M<0.1$~M$_\odot$)
the radius becomes practically independent of the mass. Some configurations of grazing binary eclipses and
of eclipsing binaries in multiple systems can also mimic a planetary transit signal.

%Spectroscopic follow-up 
%observations are required to reveal the real nature of the system. The Doppler information gives the mass of the transiting object,
%and the spectroscopic parameters indicate the size of the main star, hence the precise size of the transiting companion.

%\section{Sorting out OGLE planetary transit candidates}

Some indications on the nature of the OGLE transiting companion can be gathered from the light curve (e.g. Sirko \& Paczyincki 2003) and with low-resolution spectroscopy (e.g. Dreizler et al. 2003, Gallardo et al. 2005). 
However, high-accuracy  radial velocity follow-up is the only way to confirm the 
exact nature of the systems by measuring the true mass of the companions. 
The spectroscopy of the central star, which is a by-product 
of the radial velocity measurement, allows to constrain the radius of the star and 
hence the real size of the transiting companion. The measurement of the true mass of the companion by the radial 
velocity orbit, coupled with the measurement of its radius, also leads to a direct measurement 
of its mean density. 

The difficulty of Doppler follow-up of OGLE candidates comes from the faintness of the 
stars (with V magnitudes in the range 15-18) located in very crowded fields. 
To characterize a hot Jupiter, one needs radial velocity precision better than 
100 {\ms} and the capability to distinguish whether the system is blended by a 
third star. Radial velocity of such accuracy had never been measured before for such faint stars.

%First of all, we selected among the 137 OGLE candidates the most promising ones in terms 
%of planetary transits according to the following criteria :
%(1) the depth and duration of the transit; (2) the shape of the transit; (3) the 
%significance of detection; (4) the amplitude of the sine and double-sine modulations 
%seen in the light curve outside the transits. 

\section{Follow-up of OGLE candidates with VLT/FLAMES}

%Our group\footnote{Francois Bouchy (LAM/OHP), Nuno Santos (Lisbon), Claudio Melo (ESO), Fr\'ed\'eric Pont, Didier Queloz, Stephane Udry and Michel Mayor (Geneva)} has observed 
Sixty of the most promising OGLE candidates from the first two seasons (OGLE-TR-1 to TR-137) were observed with the FLAMES facility on the VLT (Bouchy et al. 2005, Pont et al. 2005a). 
FLAMES is a multi-fiber link which feeds the UVES echelle spectrograph with up
to 7 targets in a field-of-view of 25 arcmin diameter, in addition to a simultaneous
Thorium calibration. The fiber link allows a stable illumination at the entrance of the 
spectrograph, and the simultaneous Thorium calibration is used to track instrumental drift. 
As a result the systematics in the radial
velocity measurements are reduced to less than 35 \ms. A 45-minute exposure on a $V=17$ magnitude target gives in a photon-noise limit of 30 \ms\ on the radial velocity for late-type, slow-rotating targets. Combining photon noise and systematics, typical precisions of 40-60 {\ms} are reached on each individual Doppler measurements for the OGLE planet-host targets.
 
Radial velocity orbits of planetary amplitudes were detected with FLAMES for five of the targets (see Table~\ref{ogle}), one of them (OGLE-TR-56) already known from similar measurements by Konacki et al. (2003). 

\begin{table}[b!th]
\begin{tabular}{l l l l l }
 & Period & Transit Depth & Planet  & Planet  \\ 
  & [days] & [\%] & mass [R$_J$] & radius [M$_J$] \\ \hline
 OGLE-TR-10 & 3.10  & 1.9 & 0.57 & 1.24 \\
 OGLE-TR-56 & 1.21  & 1.3 & 1.45 & 1.23\\
 OGLE-TR-111 & 4.01 & 1.9 & 0.53 & 1.00\\
 OGLE-TR-113 & 1.43 & 3.0 & 1.35 & 1.08\\
 OGLE-TR-132 & 1.69 & 1.1 & 1.19 & 1.13\\ \hline
\end{tabular}
\caption{Data for the OGLE transiting planets. The uncertainties are $\sim$10\% on the masses and $\sim$5\% on the radii.}
\label{ogle}
\end{table}

The characteristics of the five OGLE transiting planets have already lead to a series of interesting conclusions on hot Jupiters, the most prominent being the existence of the so-called ``very hot Jupiters'' (Bouchy et al. 2004), gas giants on very short orbits (shorter than 2 days) and heavier on average (Mazeh et al. 2004) than the more common $P>3$-days hot Jupiters.

\section{Sorting out planetary transits from impostors}

Along with the five transiting hot Jupiters, the spectroscopic follow-up programmes has led to the characterisation of more than 50 cases of ``planetary transit impostors'', i.e. configurations that could mimic the photometric signal of a planetary transit within the level of photometric noise of a ground-based transit survey.
These systems fall into four categories.
Let us review these four types of impostors in terms of implications for the follow-up of transit surveys:\smallskip

\noindent
{\bf (1) Grazing eclipsing binaries}\\
Two large stars, when eclipsing at an inclined angle, can produce shallow transit-like dips in the light curve. These cases produce, on average, rather deep signals in the light curve and are the easiest to discriminate.  Several hints are usually present in the light curve itself, such as a V-shaped transit curve, ellipsoidal modulations due to tidal effects, or a mismatch between the transit duration and the transit depth assuming a planet-sized transiting body. Nevertheless, at low signal-to-noise such systems can also be mistaken for planetary transits. They are easy to resolve with spectroscopic observations, thanks to the presence of two sets of lines in the spectra with large velocity variations.\smallskip

\noindent
{\bf (2) M-dwarf transiting companions}\\
A small M-dwarf transiting a larger star can produce a photometric signal closely similar to a planetary transit. If the companion is not larger than a hot Jupiter, and the orbital distance is too large for tidal and reflection effects to be detectable in the light curve, then the photometric signal is strictly identical to that of a planetary transit. In both cases, an opaque, Jupiter-size object transits the target star. These cases can only be resolved by Doppler observations, the amplitude of the reflex motion of the star revealing the mass of the transiting companion. Two nice examples of planet-size transiting stellar companions were found in our FLAMES follow-up among the OGLE candidates, OGLE-TR-122 (Pont et al. 2005b) and OGLE-TR-123 (Pont et al. 2005c). In particular, OGLE-TR-122, with a period of 7.2 days and a companion size smaller than that of HD209458b, produces a light curve that is strictly identical to that of a planetary transit down to a very high level of detail.\smallskip

\begin{figure}[t!]
\epsfig{file=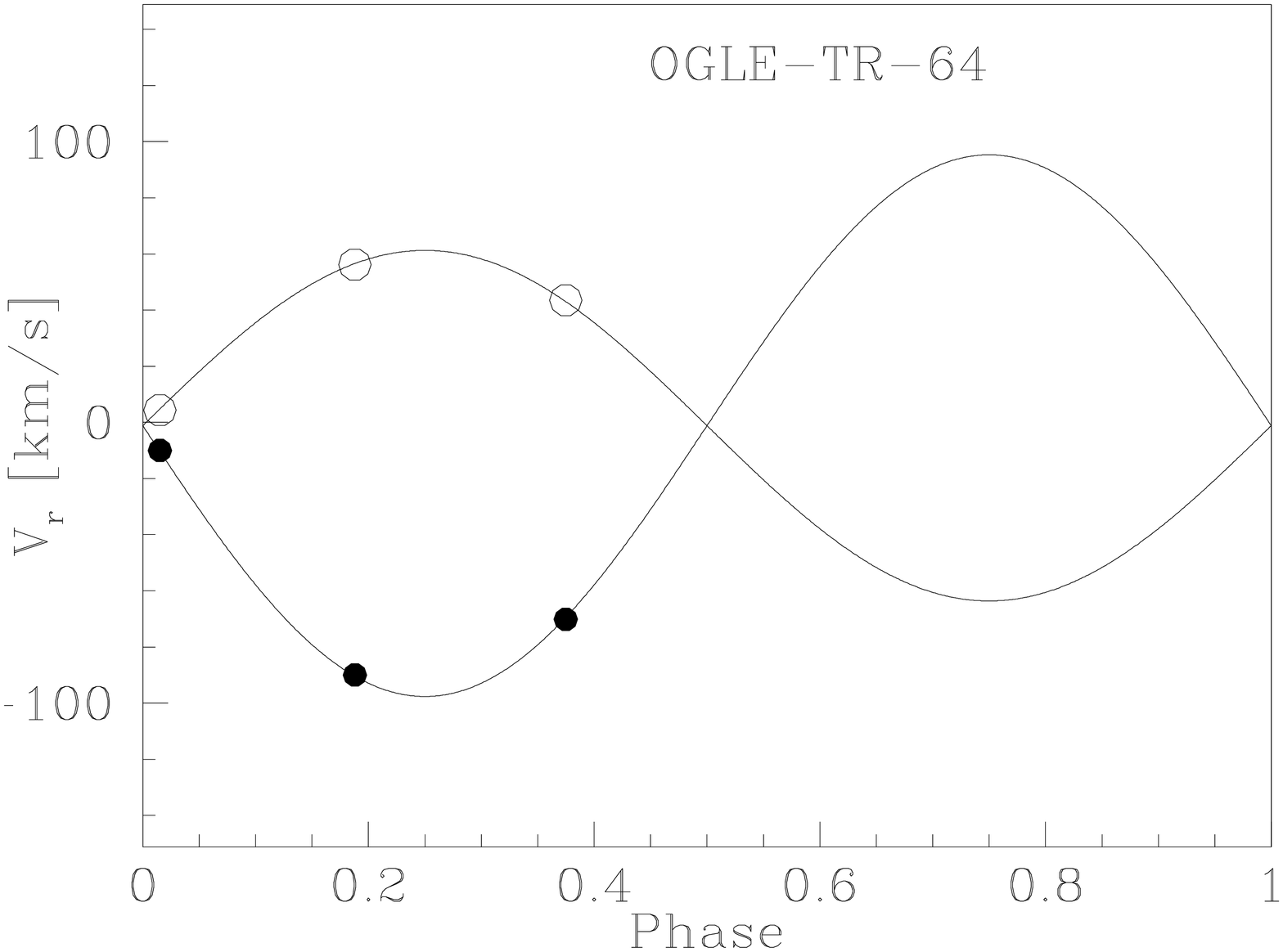,width=4.5cm,angle=-90}\epsfig{file=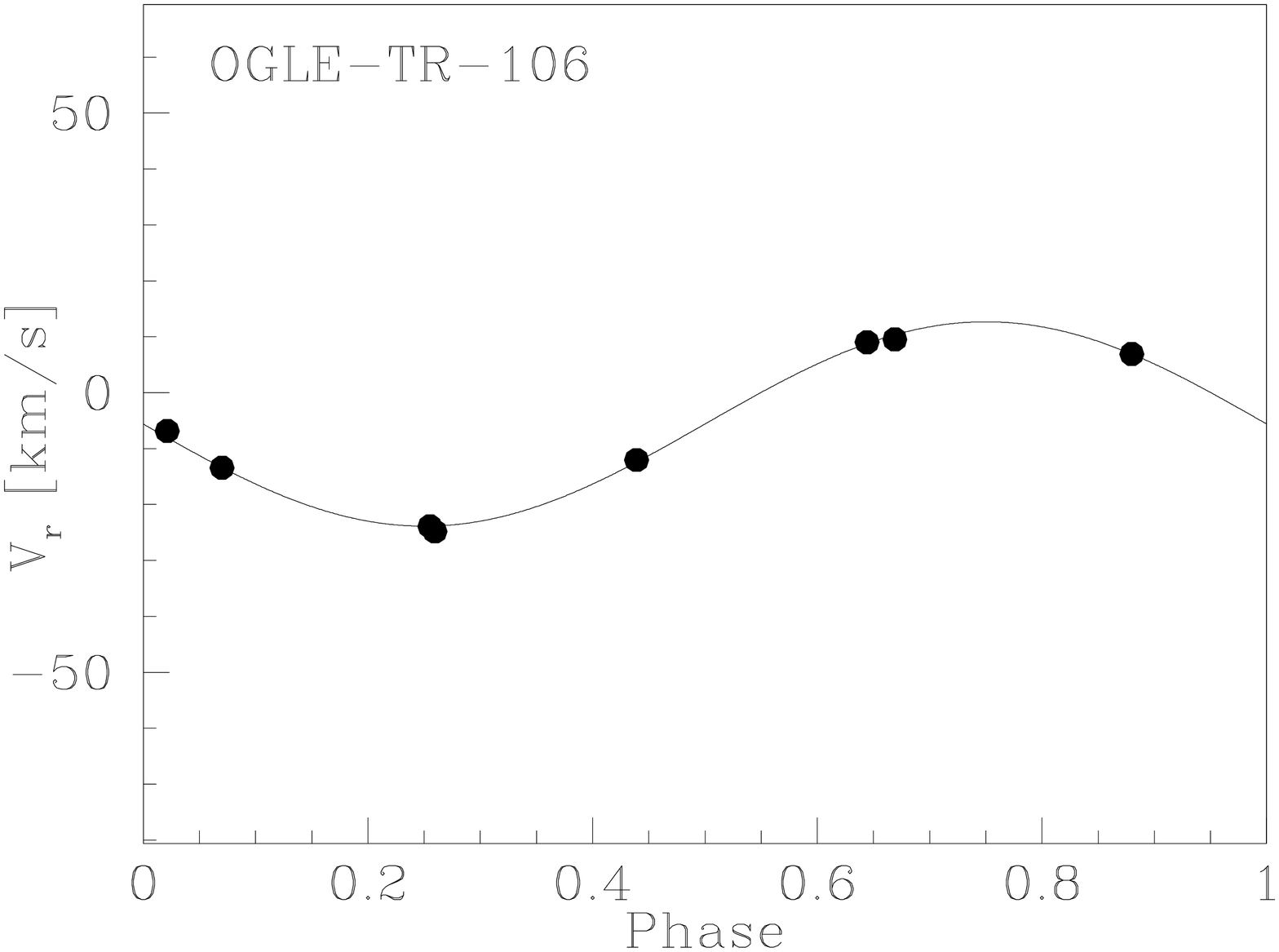,width=4.5cm,angle=-90}\\
\epsfig{file=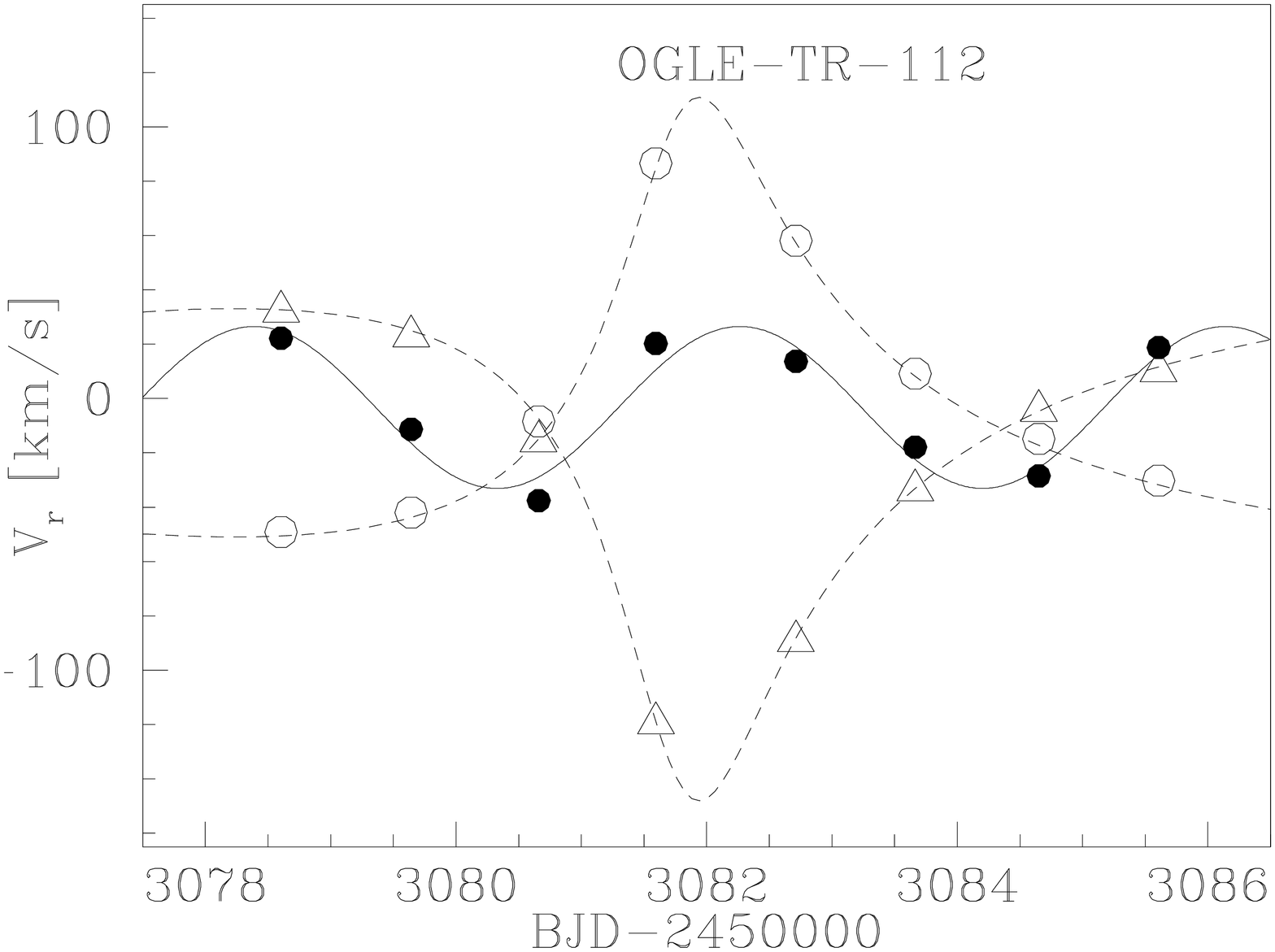,width=4.5cm,angle=-90}\epsfig{file=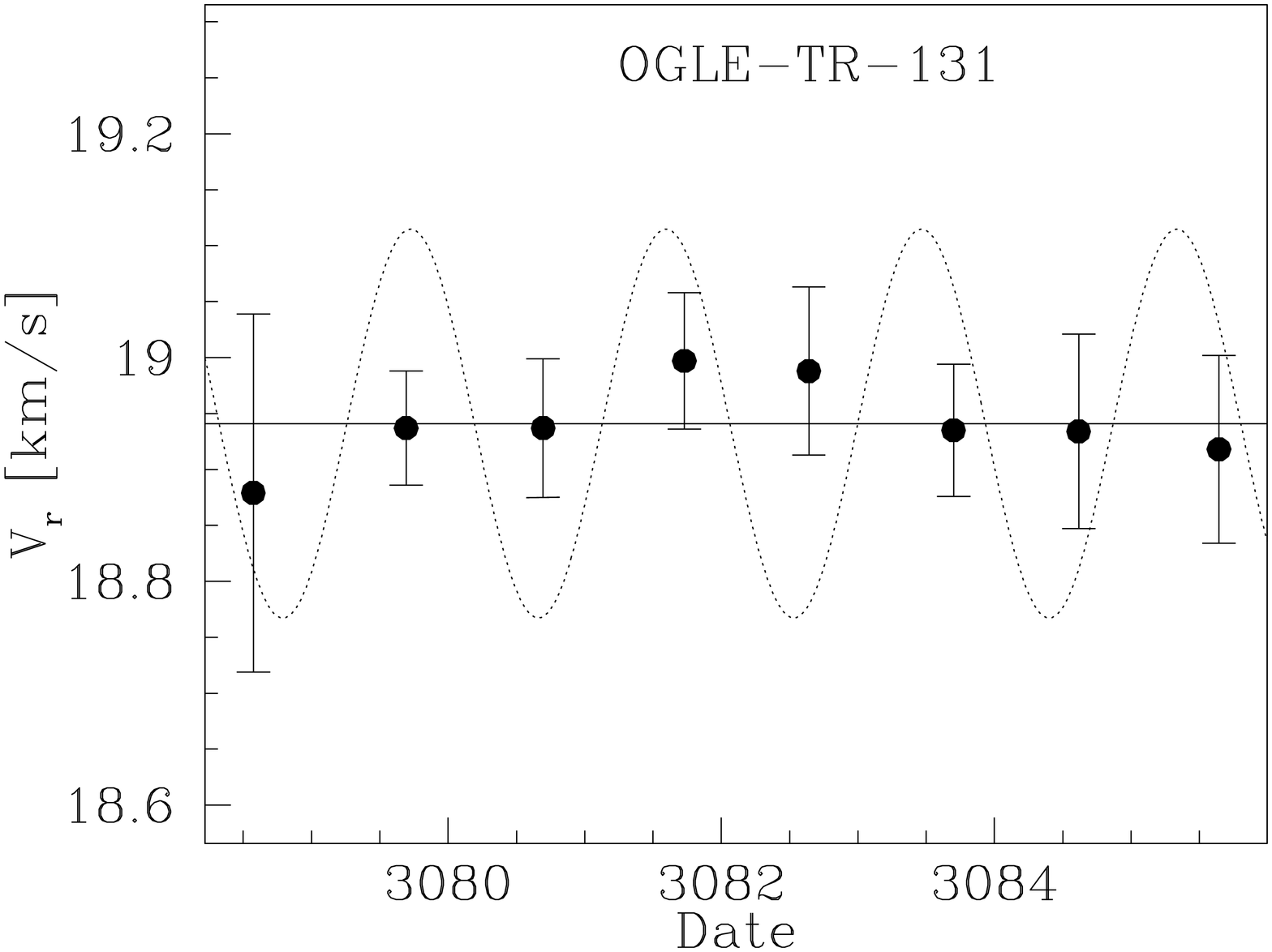,width=4.5cm,angle=-90}
%\plotone{box.ps}
\caption{Example of radial velocity data for the four types of planetary transit ``impostors'': grazing binary (top left), low-mass companion (top right), multiple system (bottom left) and false positive (bottom right, the dotted line indicates a Jupiter-mass orbit). Figure from Pont et al. (2005a).}
\label{vr}
\end{figure}

\noindent
{\bf (3) Multiple systems}\\
An eclipsing binary can produce shallow transit-like signals if the eclipse is diluted by the light of a third star. There are many possible configurations for such systems, and as a result they can be very difficult to disentangle, even with Doppler information. In most cases, multiple systems are readily discriminated with high-resolution spectroscopy from the presence of several systems of lines in the spectra (see Figure~\ref{vr}, lower left panel). However, in some cases, the parameters can conspire  not only to mimic the light curve of a planetary transit, but also to induce planet-like variations of the inferred radial velocity, produced by the blending of several sets of lines in the spectra. OGLE-TR-33 (Torres et al. 2004) is such a case. Another similar case was found in the TrES survey (Mandushev et al. 2005). \smallskip
 
\noindent
{\bf (4) False positives}\\
Stellar variability and systematic trends in the photometry can produce fluctuations in the light curve interpreted as a possible transit signal, especially as one tries to detect shallower signals near the detection threshold. OGLE-TR-58, for instance, was found to exhibit an intrisic level of variability that could explain the transit-like signal detected by OGLE without invoking a transiting companion (Bouchy et al. 2005). Further photometric observations at the epoch of the detected signal are needed in these cases to distinguish bona fide transits from false positives. \smallskip

The FLAMES/VLT Doppler follow-up, by illustrating the ubiquity of the ``impostors'' and the ability of certain configurations to mimic several aspects of the signal of bona fide transiting planets, showed that the detection of a planetary radial velocity orbit was mandatory to establish the planetary nature of a transit candidate. Moreover, the spectroscopic data must be of high-enough resolution and signal-to-noise to be able to study the evolution of the line shape (bisector analysis) over the course of the transit phase, to eliminate scenarios of multiple systems. The present practical limit for such observations is around $V=17$ mag with FLAMES on the VLT. This has 
%two
an important implication for transit surveys: 
%
%\begin{enumerate}
%\item{There can be no secure detection of transiting planets around hot stars. Although one could a priori expect transit %surveys to be a good way to circumvent the limitations of Doppler searches -- which cannot get precise radial velocity for %early-type stars because of the lack of narrow metallic line -- the necessity of the radial velocity confirmation implies %that transit surveys are subject to the same limitations in terms of spectral type of the primary.}
%\item{
deep transit surveys using large telescopes and the HST will produce candidates that are to faint to be confirmed spectroscopically with present-day telescopes ($V>17$), and therefore will not lead to any confirmed transiting planet detection! This is in contrast to their very good detection rates ``on paper'' and represents a large drawback for such programmes.
%}
%\end{enumerate}

%\section{Transit surveys 10 years after 51 Peg}
\section{Detection threshold and expected yield of transit surveys}

Three transiting exoplanets have been found by radial velocity searches, out of a few thousand target stars.  This is roughly in line with prior expectations:  the rate of occurence of hot Jupiter transits is about one in a thousand.
On the other hand, photometric transit surveys have found 1 transiting planet around a relatively bright star (TrES-1), and 5 around fainter stars (OGLE). This is much lower than initial expectations, with hundreds of thousands of field stars having been sampled by several dozen surveys ({\small STARE/WASP/VULCAN/EXPLORE/RAPTOR/PSST/SLEUTH/PISCES/ OGLEIII/STEPS/BEST/UNSW} etc...). 

The case of the most successful transit search so far, the OGLE survey, gives some precious hint as to the basic reason behind this meager results. 
%\section{Detection threshold of transit surveys in the presence of {\em real} noise}
Table~\ref{ogle} gives the period and transit depth of the five planets detected by the OGLE survey. A remarkable conclusion stems from this table: the detected planets all have exceptionally favourable parameters for transit detection. A typical hot Jupiter hosted by a typical field star will have a period above 3 days and a transit depth of the order of 1\%. By contrast, each of the detected planets has a combination of at least two of these three factors favouring detection:\smallskip\\ 
- A very short periods ($P<2$ days) -- thus providing many more transit signals in a given survey duration;\\
- A period resonant with the 1-day peak of the observation window function, like OGLE-TR-111 at almost exactly 4 days or OGLE-TR-132 near 5/3 days -- which allows the transit signal to be oversampled;\\
- A host star much smaller than the average in the field -- which causes a large transit depth.
\smallskip

The implication of this is that no ``normal'' transiting hot Jupiter was detected by the OGLE survey. 
Thus the OGLE survey -- again, the most successful ground-based survey to date -- had to rely on the very large number of targets observed to pick up a few exceptional transiting hot Jupiters. Its detection threshold was actually too high to detect normal hot Jupiters! TrES-1 also produces an exceptionally deep transit, and it is likely that our considerations apply to other ground-based transit survey as well. 

%\subsection{Theoretical detection thresholds with white noise}

%\begin{figure}[htb!]
%\epsfig{file=box.ps,width=4cm}\epsfig{file=box.ps,width=4cm}
%\plotone{box.ps}
%\caption{Estimating the significance of a transit detection.}
%\end{figure}

In theory, estimating the detection thresold and expected yield of a given transit survey is rather straightforward. The transit detection procedure is akin to finding a periodic square-shaped decrease in the flux from the target. The signal-to-noise ratio of the detection is the significance of the difference between the signal during the putative transit and the signal outside the transit.  If most data points are outside the transit, the uncertainty on the continuum level is negligible, and the detection signal-to-noise is simply the transit depth divided by its uncertainty:

\begin{equation}
SNR = \frac{depth}{\sigma / \sqrt{n}}
\end{equation}
\noindent
where $\sigma$ is the photometric uncertainty on individual points and $n$ is the number of data points during the transit.

To compute the expected yield of a given survey, one can simulate the  population of target stars, assume a frequency of planet, then compute the expected number of detections given two conditions: (1) that at least 2 or 3 transits are observed (to establish the periodicity of the signal);
(2) that the detection SNR is above some threshold, SNR$_{min}$. The SNR$_{min}$ threshold is usually assumed between 7 and 10 according to the number of false detections deemed acceptable.

Such simulations have been done for many existing and planned surveys. If the survey duration is sufficient for at least 3 transits to be observed for most hot Jupiters, then these simulations invariably predict good SNR detectability for normal hot Jupiters, at least for the brightest targets, hence resulting in significant predicted yields. 

However, there is an important hidden assumption in Eq.~1 above: it is based on the assumption of white, independent noise. If the noise is not independent and has some covariance structure, then the equivalent formula is

\begin{equation}
SNR = \frac{depth}{\sqrt{\sigma^2/n + 1/n^2 \sum_{i\neq j} cov[i;j]}}
\end{equation}

\noindent
where the $cov[i;j]$ are the elements of the covariance matrix. Therefore, the estimated yields based on the assumption of white noise are correct only if $\sigma^2/n >> 1/n^2 \sum_{i\neq j} cov[i;j]$. However, in real ground-based data in the relevant regime for hot Jupiters, the opposite is true! Plugging representative numbers shows that generally $1/n^2 \sum_{i\neq j} cov[i;j]>\sigma^2/n $. For instance in the OGLE survey, $\sigma=3-10$ mmag, $n=20-50$, so that $\sigma^2/n \simeq 0.2 - 5$ mmag$^2$, whereas the covariance of the residuals sampled on constant lightcurves gives a covariance term $1/n^2 \sum_{i\neq j} cov[i;j] \simeq  9-11$ mmag$^2$.

%\subsection{Pink noise}

\begin{figure}[t!hb]
\epsfig{file=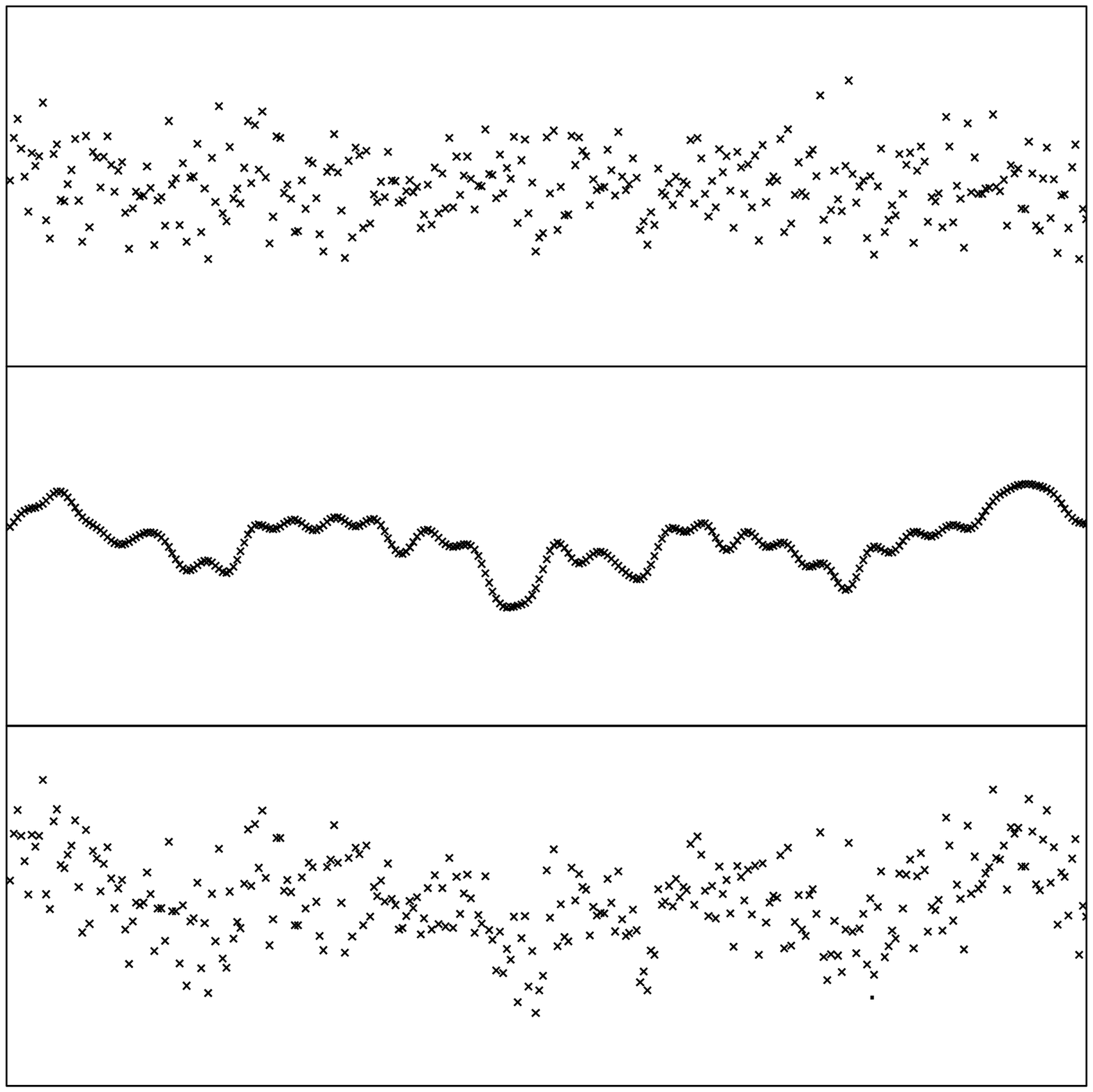,width=6cm}\epsfig{file=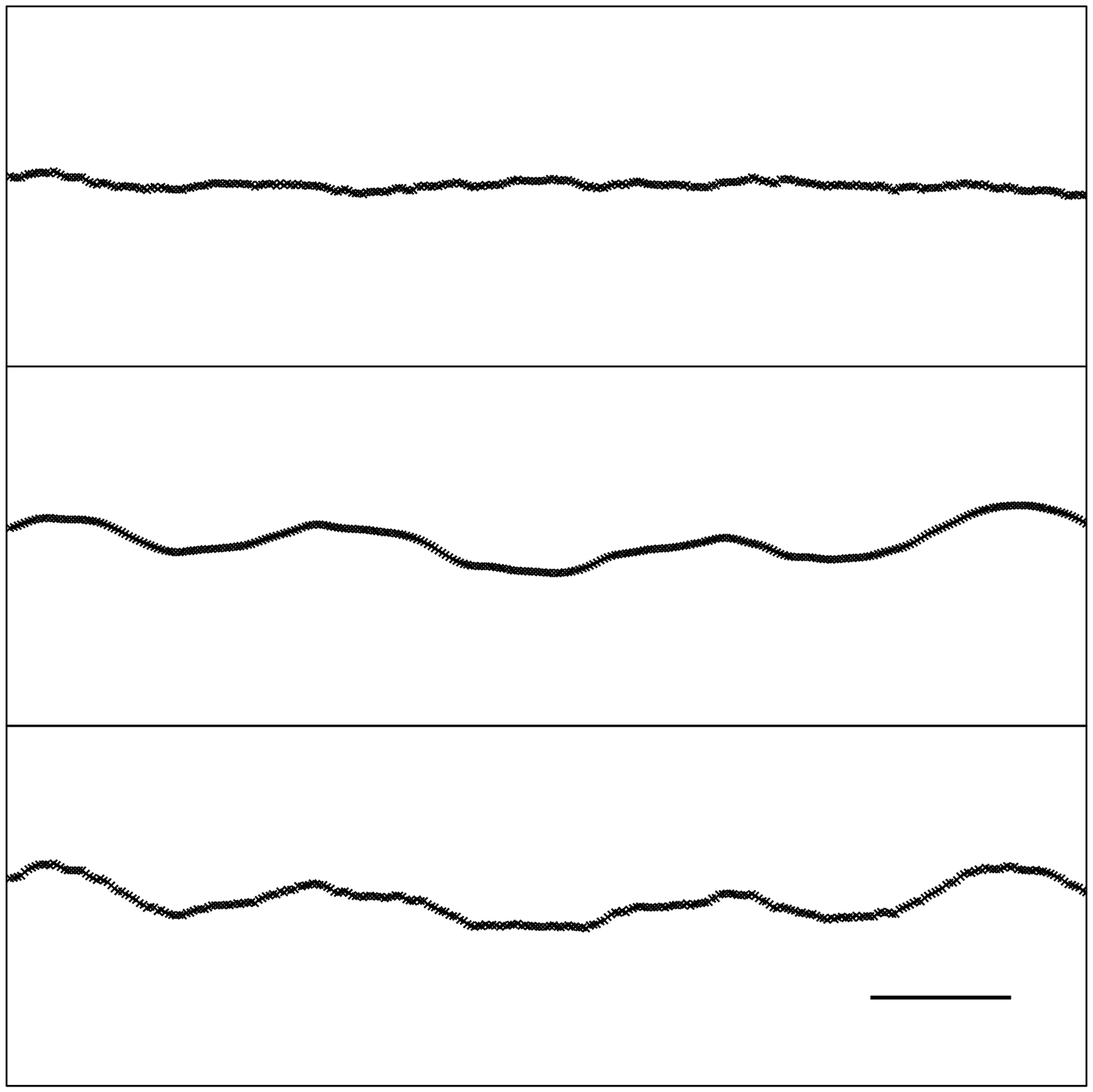,width=6cm}
%\plotone{box.ps}
\caption{{\bf Left:} A photometric time series with white, red or pink noise. The global dispersion is the same for the three curves. {\bf Right:} the same series averaged over a transit duration (the transit duration is shown by the bar at the bottom right).}
\label{pink}
\end{figure}

In the jargon of signal analysis, the noise in photometric data has a {\em white} component (mainly photon noise) and a {\em red} component. The noise on ground-based millimagnitude photometry is {\em "pink}". The red component comes from the systematics caused by the variations in atmospheric conditions, telescope tracking and detector characteristics. Figure~\ref{pink} (left panel) displays an example of these three kinds of noise, white, red and pink. Ground-based photometric data at the millimagnitude level look like the bottom curve, with some white noise superimposed on some systematics trends on longer timescale.

It is clear that the systematic trends will limit the detectability of transit signals, especially the trends operating on hour timescale -- the timescale of transits. What Eq. 2 expresses is that the detection threshold of transit surveys will depend on the average of the photon noise over a transit-length duration and the average of the covariance over this duration. The right panel of Fig.~\ref{pink} shows the average of the curves in the left panel over a transit duration. It shows graphically what was found algebraically from Eq.~2: for transit detections, the effect of the red components dominates over that of the white component (because the white component averages out to very small values over the duration of transits, whereas the red component does not).

\section{Revised yield estimates for ground-based surveys}

The implications of the presence of red noise ("systematics") in the photometric data on the expected yields of transit surveys are fundamental. In fact, in many cases a good approximation is to ignore the white noise entirely, and to base the detection threshold on the red noise only.

This profoundly modifies the predictions for the sensitivity of ground-based transit surveys. Not only the resulting detection threshold is higher than with the white-noise assumption, it also has a different dependence on period and magnitude. For instance, the presence of red noise favours the detection of very short-period transiting planets ("very hot Jupiters") compared to longer periods.

\begin{figure}[htb!]
\centering
\epsfig{file=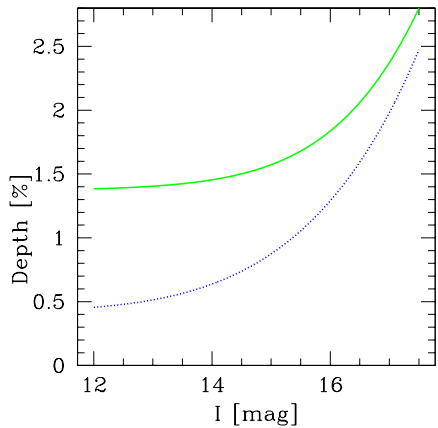,width=6cm}
%\plotone{box.ps}
\caption{Detection threshold in transit depth vs. magnitude, for a 3.5-day period planet in the OGLE survey. Dotted line: with the assumption of white noise, solid line: taking the red component of the noise into account.}
\label{thresh}
\end{figure}

Figure~\ref{thresh} gives, for the OGLE survey, the detection threshold of transit signals as a function of magnitude, assuming white noise only (dotted curve) and assuming pink noise (solid curve), for planets at $P=3.5$ days.
The difference between the two thresholds has a large effect on the expected rate of detection of hot Jupiter transits, since those are expected to produce typical transit near 1\%, exactly in the region where the prediction of white noise and pink noise diverge.

White-noise calculations predict the detectability of transit signals around the brightest stars in the survey down to very small transit depths -- thanks to the averaging of the independent noise. But taking into account the red noise leads to a much higher effective threshold, and to a floor value that is higher than the typical depth of hot Jupiter transits.
The bottom line is that the presence of systematics in the photometry drastically reduces the detectability of planetary transits in ground-based surveys.

We have devised a simplified model of the covariance matrix to be able to estimate the yields of transit surveys in the presence of red noise (Pont \& Zucker 2006, in prep.). This model uses a simplified description of the matrix based on a single parameter, called $\sigma_r$, describing the amplitude of the red noise in the relevant regime. This model was applied to some ongoing transit surveys\footnote{STARE (Brown \& Charbonneau 2000), OGLE (Udalski et al. 2002a), HAT (Bakos et al. 2004), Vulcan (Borucki et al. 2001), UNSW (Hidas et al. 2005)} estimating $\sigma_r$ from published data, to predict the yield of these surveys per season in terms of transiting hot Jupiter detections (we did not include "very hot Jupiters"). 
%The results are presented in Table~\ref{surveys}, compared to the predictions assuming white noise. 
Table~\ref{surveys} shows the results compared to the predictions assuming white noise only.

\begin{table}[h!tb]
\begin{tabular}{c |  c c c c c}
\centering
  & STARE & OGLE & HAT-5 & Vulcan & UNSW \\ \hline
HJ/season & & & & & \\
(pink noise) & 0.9 & 1.1 & 0.2 & 0.6 &0.01 \\
HJ/season & & & & & \\
(white noise) & 7 & 14 & 18 & 16 & 2 \\ \hline
\end{tabular}
\caption{Expected number of detections of transiting hot Jupiters per season for some ground-based surveys, assuming pink (correlated) noise, or white (independent) noise.}
\label{surveys}
\end{table}

%\begin{table}[htbp]
%\begin{tabular}{l l l}
%Survey name & HJ/season  & HJ/season  \\
%  & (pink noise) & (white noise) \\ \hline \hline
% STARE & 0.9 & 7\\
% OGLE & 1.1 & 14\\
% HAT-5 & 0.2 & 18\\
% Vulcan & 0.6 & 16\\
% UNSW & 0.01 & 2\\
% \end{tabular}
 %\caption{Expected number of transiting Hot Jupiter detections per season for some transit surveys. The estimates are approximate numbers based on published information, showing the effect of the pink noise (systematics in the photometry) according to our model.}
 %\label{surveys}
%\end{table}

%The results depend somewhat on the assumptions made on the covariance matrix. 
Generally, taking the red component of the noise into account results in a drastic downward revision of the expected yields, more in line with the actual rates of detection. 

Our revised estimates tend to indicate that major surveys, such as TrES or OGLE, have the potential to detect a number of transiting hot Jupiters of the order of unity per season, while more modest surveys or surveys affected by a higher level of covariance are reduced to rather negligible values. A possible implication of our studies is that, contrary to the indications of initial estimates based on white noise, transit surveys of limited scope and duration are unlikely to make a contribution to the field. Only long-term, ambitious searches with great care invested in the correction of the systematics will detect transiting hot Jupiters in significant numbers -- numbers that are going to be roughly one order of magnitude lower than white-noise estimates.

Some authors have predicted that with large telescope, transiting planets of much smaller size than hot Jupiters (``hot Neptunes'') will become detectable from the ground (Gillon et al. 2005, Hartman et al. 2005). However, the reasoning used to reach these conclusions are based on the same white-noise assumptions as those leading to the predictions of very high rates of hot Jupiter detections by the on-going surveys. When the red component of the noise is taken into account, we find that hot Neptunes are not likely to be detected in significant numbers from the ground, and that space missions like Corot and Kepler will be needed to avoid the type of hour-timescale red noise that the Earth's atmosphere is causing in light curves.

%\section{Conclusions}
%Your conclusions

%\acknowledgments{
%The editors of the proceedings want to thank very much all the 
%participants of the colloquium who will send their paper before 
%the deadline of 30th of September 2005. 
%}

\end{document}